\documentclass[preprint,showpacs,preprintnumbers,amsmath,amssymb]{revtex4}

\usepackage{graphicx}% Include figure files
\usepackage{dcolumn}% Align table columns on decimal point
\usepackage{bm}% bold math
\usepackage{longtable}

\begin{document}

\title{Candidate chiral doublet bands in the odd-odd nucleus $^{126}$Cs}% Force line breaks with \\
\author{Shouyu Wang,$^{1}$\  Yunzuo Liu,$^{1, 2}$  T. Komatsubara,$^{3}$ Yingjun Ma,$^{1}$ and Yuhu Zhang$^{2}$\\
 $^1$Department of Physics, Jilin University, Changchun 130021, P. R. China\\
 $^2$Institute of Modern Physics, Chinese Academy of Sciences, Lanzhou 730000, P. R. China\\
 $^3$Institute of Physics and Tandem Accelerator Center, University of Tsukuba, Ibaraki 305, Japan\\
}

\begin{abstract}
The  candidate chiral doublet bands recently observed in
$^{126}$Cs have been extended to higher spins, several new linking
transitions between the two partner members of the chiral doublet
bands are observed, and $\gamma$$-$intensities related to the
chiral doublet bands are presented by analyzing the
$\gamma$$-$$\gamma$ coincidence data collected earlier at the
NORDBALL through the $^{116}$Cd$($$^{14}$N, 4n$)$$^{126}$Cs
reaction at a beam energy of 65 MeV. The intraband $B(M1)/B(E2)$
and interband $B(M1)_{in}/B(M1)_{out}$ ratios and the energy
staggering parameter, S(I), have been deduced for these doublet
bands. The results are found to be consistent with the chiral
interpretation for the two structures. Furthermore, the
observation of chiral doublet bands in $^{126}$Cs together with
those in $^{124}$Cs, $^{128}$Cs, $^{130}$Cs and $^{132}$Cs also
indicates that the chiral conditions do not change rapidly with
decreasing neutron number in these odd-odd Cesium isotopes.
\end{abstract}

\maketitle

Chiral doublet bands have been predicted~\cite{FM97} to appear in
odd-odd nuclei with substantial triaxial deformation and with
configuration where valence proton lies in the lower part of the
$\pi h_{11/2}$ subshell and valence neutron lies in the higher
part of the $\nu h_{11/2}$ subshell in the $A\sim130$ mass region.
Extensive experimental studies have been made to search for such
bands and, as results, candidate chiral doublet bands have been
reported in several N=73~\cite{Koike01}, N=75~\cite{Starosta01}
and N=77~\cite{GR03,Bark01} isotones.

Even-even nuclei with A$\sim$130 are known to be $\gamma$-soft and
under the influences of prolate driving, particle-like, $h_{11/2}$
proton orbital and oblate driving, hole-like, $h_{11/2}$ neutron
orbital, the odd-odd nuclei with A$\sim$130 are expected to have
relatively stable triaxial shapes. For odd-odd Cs isotopes with
A$\sim$130, the proton Fermi level lies in the bottom of the $\pi
h_{11/2}$ subshell and the neutron Fermi level approaches the top
of the $\nu h_{11/2}$ subshell with increasing neutron number N.
Odd-odd Cs isotopes with larger N are more suitable for sustaining
the chiral geometry than those with smaller N and thus odd-odd Cs
isotopes with larger N are better candidates for the observation
of chiral doublet bands unless the value of N is too close to N=82
where the rotational structures are not to be developed. Chiral
doublet bands have been reported in $^{128}$Cs~\cite{Koike03},
$^{130}$Cs~\cite{AJ05} and $^{132}$Cs~\cite{GR03}. On the other
band, two positive-parity bands had previously been reported in
$^{124}$Cs~\cite{AG01}, but in this case they were not interpreted
in terms of the chiral rotation. Recently, the data of the two
positive-parity bands in $^{124}$Cs, reported in Ref~\cite{AG01},
were systematically compared to those of the chiral doublet bands
in $^{128}$Cs and $^{130}$Cs by Koike et al.~\cite{Koike03} and it
was found that the degree of degeneracy and the intraband
$B(M1)/B(E2)$ and interband $B(M1)_{in}/B(M1)_{out}$ ratios of the
two positive-parity bands in $^{124}$Cs are similar to those of
the chiral doublet bands in $^{128,130}$Cs~\cite{Koike03}. This
report presents the extended level scheme and $\gamma$-intensities
of the candidate chiral doublet bands in $^{126}$Cs, and completes
the systematic comparison of the electromagnetic properties of the
candidate chiral doublet bands in the series of odd-odd Cs
isotopes from N$=$69 to N$=$77.

Attempting to search for the chiral doublet bands in $^{126}$Cs,
high-spin states of $^{126}$Cs were investigated by Li et
al.\cite{LiXF02} and candidate chiral doublet bands in $^{126}$Cs
were proposed. However, due to poorer counting statistics,
electromagnetic properties of the candidate chiral doublet bands
were not discussed in Ref.\cite{LiXF02}. The high-spin states of
$^{126}$Cs had previously been studied by T. Komatsubara et
al.~\cite{TK93} where the coincidence measurements were performed
at the NORDBALL, but $\gamma$$-$intensities were not provided in
Ref. \cite{TK93}. It is expected that the counting statistics of
$\gamma$$-$$\gamma$ coincidence measurements in Ref.~\cite{TK93}
could be much better than that of Ref.~\cite{LiXF02}. In order to
study electromagnetic properties of the chiral doublet bands in
$^{126}$Cs, we reanalyzed the $\gamma$$-$$\gamma$ coincidence data
collected earlier at NORDBALL by Komatsubara et al.\cite{TK93}. In
the present reanalysis of the experimental data collected in Ref.
\cite{TK93}, the coincidence data were sorted into a 4096 by 4096
channel symmetrized $E_{\gamma}-E_{\gamma}$ matrix. To obtain
information on the $\gamma-$ray multipolarities, two asymmetric
matrices were constructed and then ADO ratios ( $\gamma-$ray
angular distribution from oriented nuclei) were evaluated using
the method as described in Ref. \cite{MP96}. Typical ADO ratios
observed for the known $\gamma-$rays were 1.4 for stretched
quadrupole or $\Delta$I=0 dipole radiations and 0.7 for stretched
dipole ones. This empirical law can thus be used to assist us in
the multipolarity assignments for newly observed $\gamma-$rays.
This data set had been utilized previously to establish the high
spin structure of $^{123}$I \cite{WANG06}. The detailed
experimental setup and procedure were described in Ref.
\cite{WANG06}. $\gamma-$rays assigned to $^{126}$Cs are listed in
Table I, together with their $\gamma$$-$intensities, ADO ratios,
and spin-parity assignments.

Partial level scheme for $^{126}$Cs derived from the present work
is shown in Fig. 1, where band 1 is the yrast band and band 2 is
the side band, and the bandhead spin $I_{0}=9$ of the yrast band
was adopted from Ref. \cite{YLiu98}. The previously reported
positive-parity doublet bands have been extended, for the side
band, the odd-spin decay sequence has been extended from $21^{+}$
to $23^{+}$ and even-spin decay sequence from $16^{+}$ to
$22^{+}$, and linking transitions 315.5, 739.0, 785.0, 791.8,
847.8, 937.8 and 1100.0 keV between the two bands have been added
to the level scheme. Sample spectra are shown in Fig. 2. The new
observation of both M1 and E2 linking transitions between the two
bands provides additional support to the judgement that the side
band has the same parity as that of the yrast band. The alignment
plots of band 1, band 2 and the known $\pi h_{11/2}\otimes\nu
g_{7/2}$ band, reported in Ref. \cite{LiXF03}, of $^{126}$Cs  are
shown in Fig.3 where the $\pi h_{11/2}\otimes\nu g_{7/2}$ band
exhibits a sharp backbend at $\hbar\omega=$0.41$MeV$ caused by the
rotational alignment of the first pair of $h_{11/2}$ neutrons. The
absence of a band crossing at this frequency for bands 1 and 2 in
Fig. 3 indicates that the $\nu h_{11/2}$ orbital is Pauli blocked
in both bands, which suggests that an $\nu h_{11/2}$ orbital is
involved in the configurations of these bands. Moreover, the large
initial alignments ($\sim$ 6 $\hbar$) for both bands 1 and 2
strongly suggests that the $h_{11/2}$ proton is involved in the
configuration of bands 1 and 2.

All these experimental observations suggest that the side band has
the same configuration $\pi h_{11/2}\otimes\nu h_{11/2}$ as that
of the yrast band. A possible interpretation of the side band is
that it may result from the coupling between the unfavoured
signature of the $\pi h_{11/2}$ orbital and the two signatures of
the $\nu h_{11/2}$ orbital. However, an energy splitting of $\sim$
200 keV(see Fig. 4(a)) between the two bands is too small for the
side band to be interpreted as a band built on the unfavoured
signature of the proton orbital. An interpretation of
$\gamma$-vibration coupled to the yrast band can also be ruled out
because the $\gamma$-vibration energies are $\geq$ 600 keV in this
mass region~\cite{Starosta01,Koike03}.

The observation of two near-degenerate $\Delta$I=1 bands has been
considered as the fingerprint of the existence of chiral rotation.
Recently, two further fingerprints, which must be observed in
order to be consistent with the chiral geometry interpretation,
were proposed by Koike et al.~\cite{Koike02}. Firstly, the energy
staggering parameter $S (I) = [E (I) - E (I -1)] / 2I$ should
possess a smooth dependence with spin since the particle and hole
orbital angular momentum are both perpendicular to the core
rotation in the chiral geometry. Secondly, due to the restoration
of the chiral symmetry in the laboratory frame there are phase
consequences for the chiral wavefunctions resulting in $M1$ and
$E2$ selection rules which can manifest as $B(M1)/B(E2)$ and
$B(M1)_{in}/B(M1)_{out}$ staggering as a function of spin and the
odd spin members of the chiral bands have higher values relative
to the even spin members for chiral bands built on the
configurations $\pi h_{11/2}\otimes\nu h_{11/2}$ in nuclei with
$A\sim130$.

The degree of degeneracy of the two positive parity bands in
$^{126}$Cs is indicated by the excitation energies of the band
members as a function of spin as shown in Fig. 4(a). The two
curves maintain a roughly constant energy difference of $\sim200$
keV showing near degeneracy within the observed spin interval. A
plot of S(I) vs. spin I is displayed in Fig. 4(b) where no
significant spin dependence with spin is observed for I$>14\hbar$.
The reduced transition probability ratios $B(M1)/B(E2)$ and
$B(M1)_{in}/B(M1)_{out}$, which were deduced from the
$\gamma$$-$intensities listed in Table I by using the relations
(7) and (8) of Ref. \cite{Koike03}, respectively, are shown in
Fig. 5. The staggering of $B(M1)/B(E2)$ and
$B(M1)_{in}/B(M1)_{out}$ ratios is evident both for the yrast and
side band, and the odd-spin states have higher values than
even-spin states except that the $B(M1)/B(E2)$ staggering of the
side band is less pronounced and it deviates from the staggering
pattern at I=14$\hbar$. The features of the two positive parity
bands in $^{126}$Cs, displayed in Fig. 4 and Fig. 5, generally fit
to the fingerprints of chiral rotation and therefore the two
positive parity bands in $^{126}$Cs should be considered as
candidate chiral doublet bands.

For comparison, the $B(M1)/B(E2)$ and $B(M1)_{in}/B(M1)_{out}$
ratios of the two positive parity bands in $^{124}$Cs~\cite{AG01},
$^{128}$Cs~\cite{Koike03}, $^{130}$Cs~\cite{AJ05} and
$^{132}$Cs~\cite{GR03} are also displayed in Fig. 5. It is
interesting to note that the similarities of the $B(M1)/B(E2)$ and
$B(M1)_{in}/B(M1)_{out}$ ratios of these odd-odd Cs isotopes
indicate that the chiral conditions do not change rapidly with
neutron number N decreasing from N=77 ($^{132}$Cs) to N=69
($^{124}$Cs). This is difficult to understand in view of the fact
that the neutron Fermi level in $^{124}$Cs and $^{126}$Cs lies in
or close to the middle of the $\nu h_{11/2}$ subshell, where the
orthogonal coupling between the proton and the neutron angular
momentum, required by the chiral geometry, is not expected to
occur.

In summary, high-spin states in odd-odd $^{126}$Cs have been
studied by using the $^{116}$Cd($^{14}$N, 4n)$^{126}$Cs reaction.
The previously reported two positive-parity bands have been
confirmed and extended and new linking transitions between the two
bands have been observed. The properties of the two
positive-parity bands show general agreement with the fingerprints
of chiral rotation and thus these two bands are suggested to be
the candidate chiral doublet bands in $^{126}$Cs. The
$B(M1)/B(E2)$ and $B(M1)_{in}/B(M1)_{out}$ ratios deduced from the
present work show similar staggering pattern as those in
$^{124,128,130,132}$Cs. The observation of chiral doublet bands in
$^{126}$Cs indicates that the chiral conditions do not change
rapidly with decreasing neutron number in these odd-odd Cs
isotopes.

This study was supported by the National Natural Science
Foundation (Grant Nos.10275028, 10205006 and 10105003), Research
Fund for the doctoral Program of Higher Education (No.20030183055)
and the Major State Research Development Programme (No.
G2000077405) of China.

\vspace{1.0cm}
\renewcommand\refname{References}
\renewcommand{\baselinestretch}{1.5}

%\newpage
\renewcommand{\thefootnote}{\alph{footnote}}
\begin{longtable}{cccc}
\caption {Energies, $\gamma-$rays and ADO ratios of $\gamma-$rays
in $^{126}$Cs.}
\label{tab:table1} \\
\hline\hline \text{$E_{\gamma}$ (keV)} & \text{$I_{\gamma}$(\%)}
& \text{ADO Ratio} & \text{$I^{\pi}_{i}$$\longrightarrow$$I^{\pi}_{f}$}\\
\endfirsthead
\multicolumn{2}{c}{TABLE I. (Continued)}\\
\hline\hline \text{$E_{\gamma}$ (keV)} & \text{$I_{\gamma}$(\%)}
& \text{ADO Ratio} & \text{$I^{\pi}_{i}$$\longrightarrow$$I^{\pi}_{f}$}\\
\hline
\endhead
\hline
\endfoot
\hline\hline
\endlastfoot
\hline
Band 1& & &\\
140.8 & 100.0(3.0) & 0.93(0.06) & $(10^{+})$$\longrightarrow$$(9^{+})$\\
254.8 & 63.0(6.5) & 0.70(0.08) & $(12^{+})$$\longrightarrow$$(11^{+})$\\
337.5 & 62.2(3.8) & 0.65(0.09) & $(11^{+})$$\longrightarrow$$(10^{+})$\\
343.2 & 17.8(2.6) & 0.74(0.11) & $(14^{+})$$\longrightarrow$$(13^{+})$\\
395.7 & 47.9(3.2) & 0.66(0.08) & $(13^{+})$$\longrightarrow$$(12^{+})$\\
415.5 & 2.5(0.4) &  & $(16^{+})$$\longrightarrow$$(15^{+})$\\
463.5 & 26.7(2.5) & 0.66(0.08) & $(15^{+})$$\longrightarrow$$(14^{+})$\\
478.1 & 2.7(0.5) & & $(11^{+})$$\longrightarrow$$(9^{+})$\\
495.8 & 15.6(3.3) & 0.68(0.13) & $(17^{+})$$\longrightarrow$$(16^{+})$\\
527.7 & 3.0(0.7) &  & $(19^{+})$$\longrightarrow$$(18^{+})$\\
554.5 & 3.2(0.7) &  & $(21^{+})$$\longrightarrow$$(20^{+})$\\
592.5 & 53.7(5.8) & 1.36(0.19) & $(12^{+})$$\longrightarrow$$(10^{+})$\\
650.5 & 10.6(1.6) & 1.44(0.35) & $(13^{+})$$\longrightarrow$$(11^{+})$\\
739.2 & 48.2(5.5) & 1.33(0.21) & $(14^{+})$$\longrightarrow$$(12^{+})$\\
807.0 & 12.6(1.6) & 1.34(0.27) & $(15^{+})$$\longrightarrow$$(13^{+})$\\
879.0 & 32.7(1.7) & 1.43(0.25) & $(16^{+})$$\longrightarrow$$(14^{+})$\\
910.8 & 13.0(2.9) & 1.39(0.27) & $(17^{+})$$\longrightarrow$$(15^{+})$\\
960.7 & 16.5(3.5) & 1.45(0.28) & $(18^{+})$$\longrightarrow$$(16^{+})$\\
993.0 & 4.7(1.0) & 1.46(0.46) & $(19^{+})$$\longrightarrow$$(17^{+})$\\
1045.5 & 14.1(2.6) & 1.63(0.34) & $(20^{+})$$\longrightarrow$$(18^{+})$\\
1072.2 & 5.7(1.3) & 1.27(0.35) & $(21^{+})$$\longrightarrow$$(19^{+})$\\
1135.2 & 2.6(0.8) &  & $(22^{+})$$\longrightarrow$$(20^{+})$\\
1148.0 & 1.0(0.3) &  & $(25^{+})$$\longrightarrow$$(23^{+})$\\
1148.3 & 1.4(0.4) &  & $(23^{+})$$\longrightarrow$$(21^{+})$\\
1198.5 & 1.0(0.3) &  & $(24^{+})$$\longrightarrow$$(22^{+})$\\
Band 2 &  &  & \\
327.1 & 4.6(0.6) & 0.64(0.19) & $(13^{+})$$\longrightarrow$$(12^{+})$\\
344.4 & 6.4(1.0) & 0.71(0.22) & $(14^{+})$$\longrightarrow$$(13^{+})$\\
361.9 & 4.7(0.8) & 0.59(0.17) & $(12^{+})$$\longrightarrow$$(11^{+})$\\
426.5 & 11.7(1.5) & 0.66(0.16) & $(15^{+})$$\longrightarrow$$(14^{+})$\\
461.1 & 3.2(0.7) &  & $(17^{+})$$\longrightarrow$$(16^{+})$\\
671.1 & 5.7(0.8) & 1.35(0.48) & $(14^{+})$$\longrightarrow$$(12^{+})$\\
689.3 & 11.6(1.0) & 1.37(0.28) & $(13^{+})$$\longrightarrow$$(11^{+})$\\
771.0 & 9.4(1.2) & 1.36(0.34) & $(15^{+})$$\longrightarrow$$(13^{+})$\\
901.5 & 3.6(0.7) & 1.44(0.48) & $(16^{+})$$\longrightarrow$$(14^{+})$\\
936.1 & 12.3(2.3) & 1.41(0.27) & $(17^{+})$$\longrightarrow$$(15^{+})$\\
1007.6 & 6.2(1.5) & 1.48(0.45) & $(19^{+})$$\longrightarrow$$(17^{+})$\\
1013.3 & 1.6(0.5) &  & $(18^{+})$$\longrightarrow$$(16^{+})$\\
1046.5 & 0.7(0.2) &  & $(20^{+})$$\longrightarrow$$(18^{+})$\\
1064.2 & 0.6(0.2) &  & $(22^{+})$$\longrightarrow$$(20^{+})$\\
1066.0 & 4.2(1.2) & 1.51(0.46) & $(21^{+})$$\longrightarrow$$(19^{+})$\\
1106.8 & 1.7(0.4) &  & $(23^{+})$$\longrightarrow$$(21^{+})$\\
Linking Transitions & & & \\
95.5 & 5.0(1.0) & 0.94(0.26) & $(12^{+})$$\longrightarrow$$(11^{+})$\\
145.4 & 3.7(0.6) & 0.85(0.25) & $(14^{+})$$\longrightarrow$$(13^{+})$\\
253.0 & 12.5(1.1) & 0.69(0.13) & $(16^{+})$$\longrightarrow$$(15^{+})$\\
277.5 & 4.1(1.1) & 0.66(0.19) & $(18^{+})$$\longrightarrow$$(17^{+})$\\
315.5 & 2.2(0.5) &  & $(20^{+})$$\longrightarrow$$(19^{+})$\\
496.5 & 27.9(3.2) & 0.59(0.08) & $(11^{+})$$\longrightarrow$$(10^{+})$\\
521.1 & 14.9(1.1) & 0.72(0.08) & $(12^{+})$$\longrightarrow$$(11^{+})$\\
542.5 & 25.1(2.5) & 0.64(0.06) & $(14^{+})$$\longrightarrow$$(13^{+})$\\
593.5 & 5.3(0.9) & 0.74(0.2) & $(13^{+})$$\longrightarrow$$(12^{+})$\\
637.0 & 8.4(2.5) & 0.66(0.14) & $(16^{+})$$\longrightarrow$$(15^{+})$\\
637.5 & 9.4(2.6) & 1.38(0.31) & $(11^{+})$$\longrightarrow$$(9^{+})$\\
679.8 & 2.1(0.6) &  & $(16^{+})$$\longrightarrow$$(14^{+})$\\
739.0 & 6.4(1.5) & 0.77(0.25) & $(18^{+})$$\longrightarrow$$(17^{+})$\\
785.0 &  &  & $(22^{+})$$\longrightarrow$$(21^{+})$\\
791.8 & 3.8(1.0) & 0.72(0.21) & $(20^{+})$$\longrightarrow$$(19^{+})$\\
847.8 & 2.7(0.9) &  & $(13^{+})$$\longrightarrow$$(11^{+})$\\
937.8 & 1.4(0.4) &  & $(14^{+})$$\longrightarrow$$(12^{+})$\\
968.9 & 7.0(2.1) & 1.34(0.37) & $(15^{+})$$\longrightarrow$$(13^{+})$\\
1100.0 & 1.4(0.4) &  & $(16^{+})$$\longrightarrow$$(14^{+})$
\end{longtable}

\begin{figure}[h!]
\includegraphics[bb=110 30 779 623,scale=0.5]{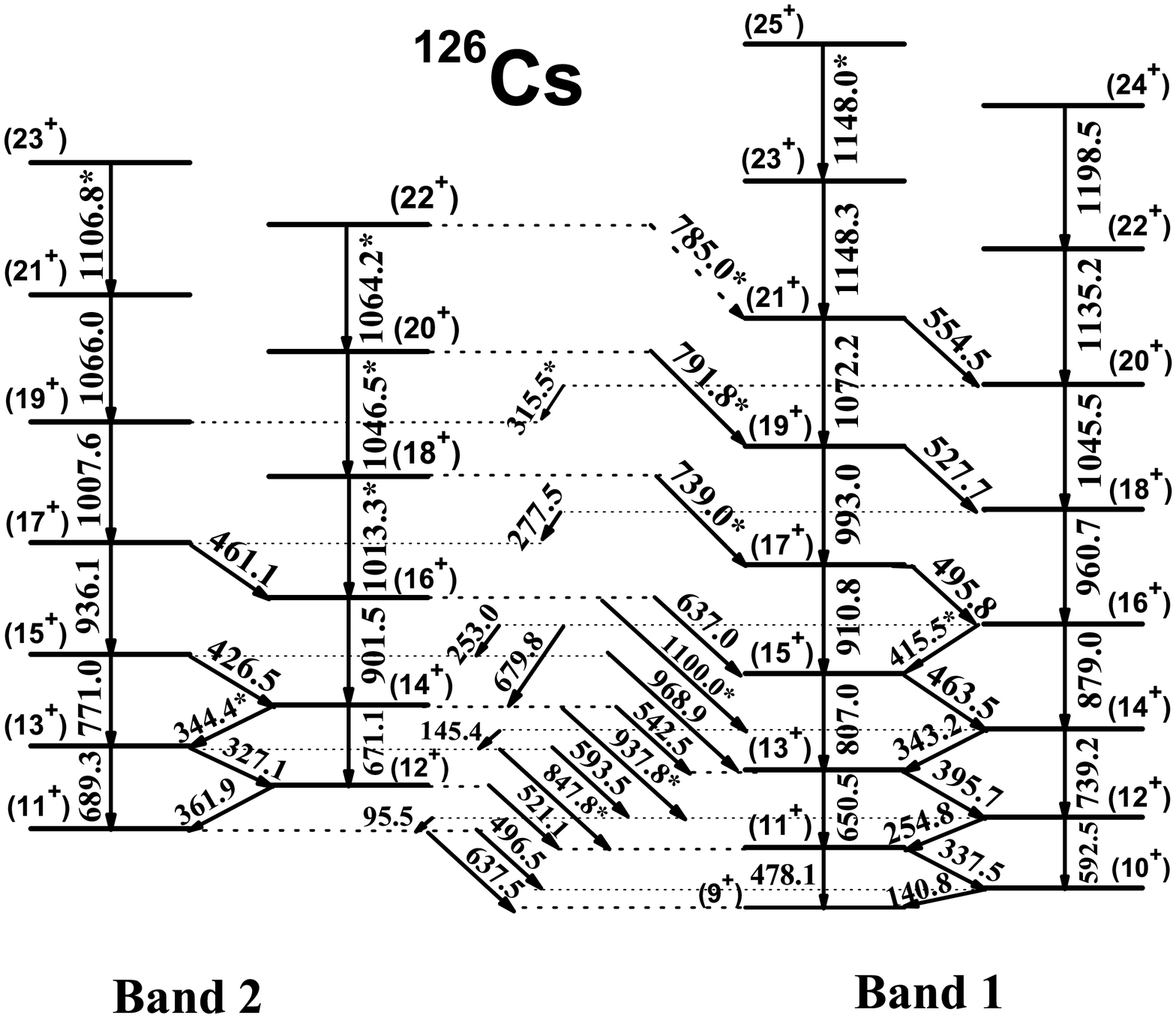}% Here is how to import EPS art
\caption{\label{fig:epsart} Partial level scheme of $^{126}$Cs.
New transitions observed in the present work are indicated with a
star.}
\end{figure}

\begin{figure}[h!]
\includegraphics[bb=50 35 645 445,scale=0.5]{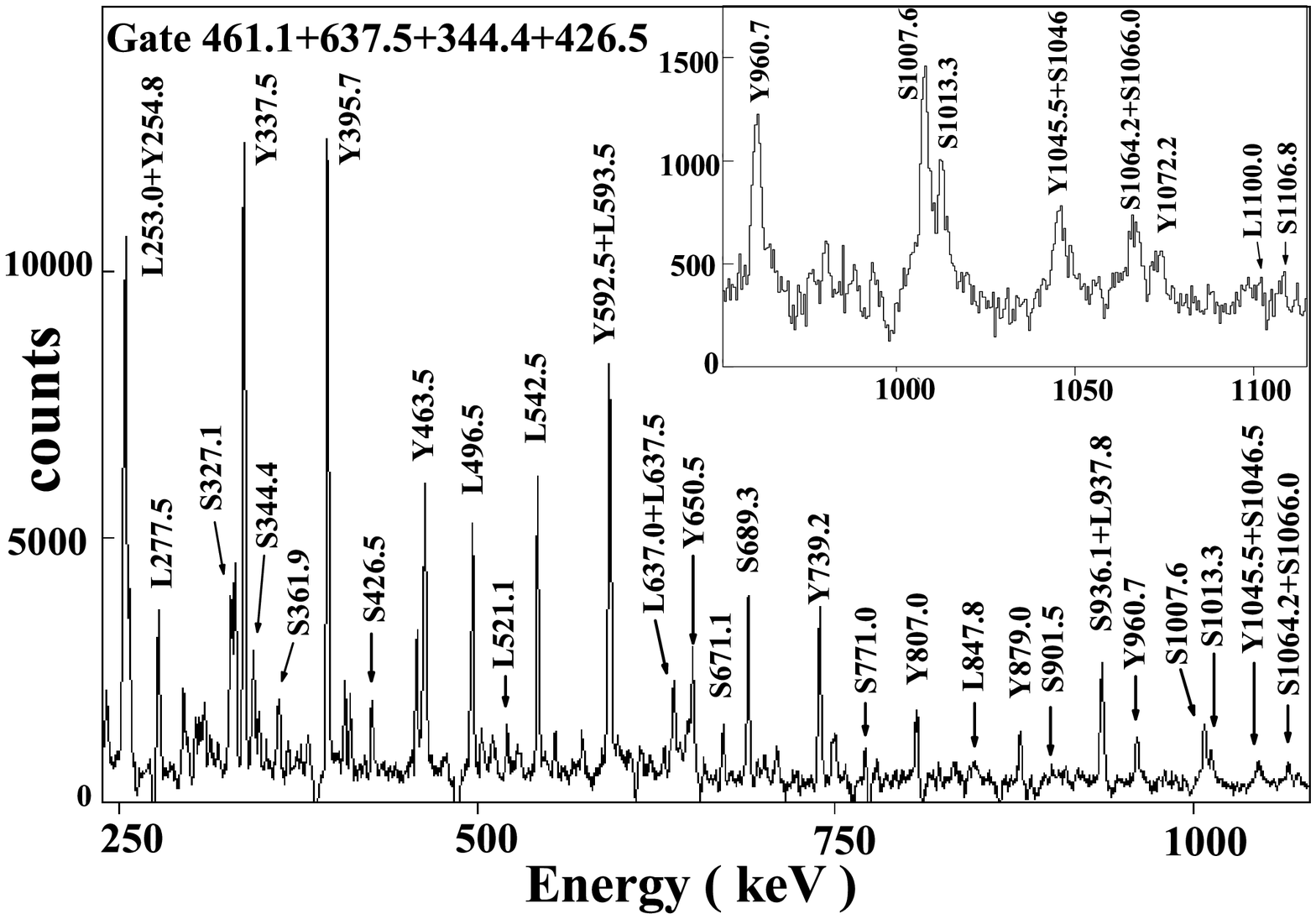}% Here is how to import EPS art
\caption{\label{fig:epsart} Sample $\gamma-$ray coincidence
spectra of the chiral doublet bands in $^{126}$Cs. Y, S, and L
stands for transitions in the yrast band, the side band, and the
linking transitions between the two bands, respectively.}
\end{figure}

\begin{figure}[h!]
\includegraphics[bb=0 0 680 530,scale=0.40]{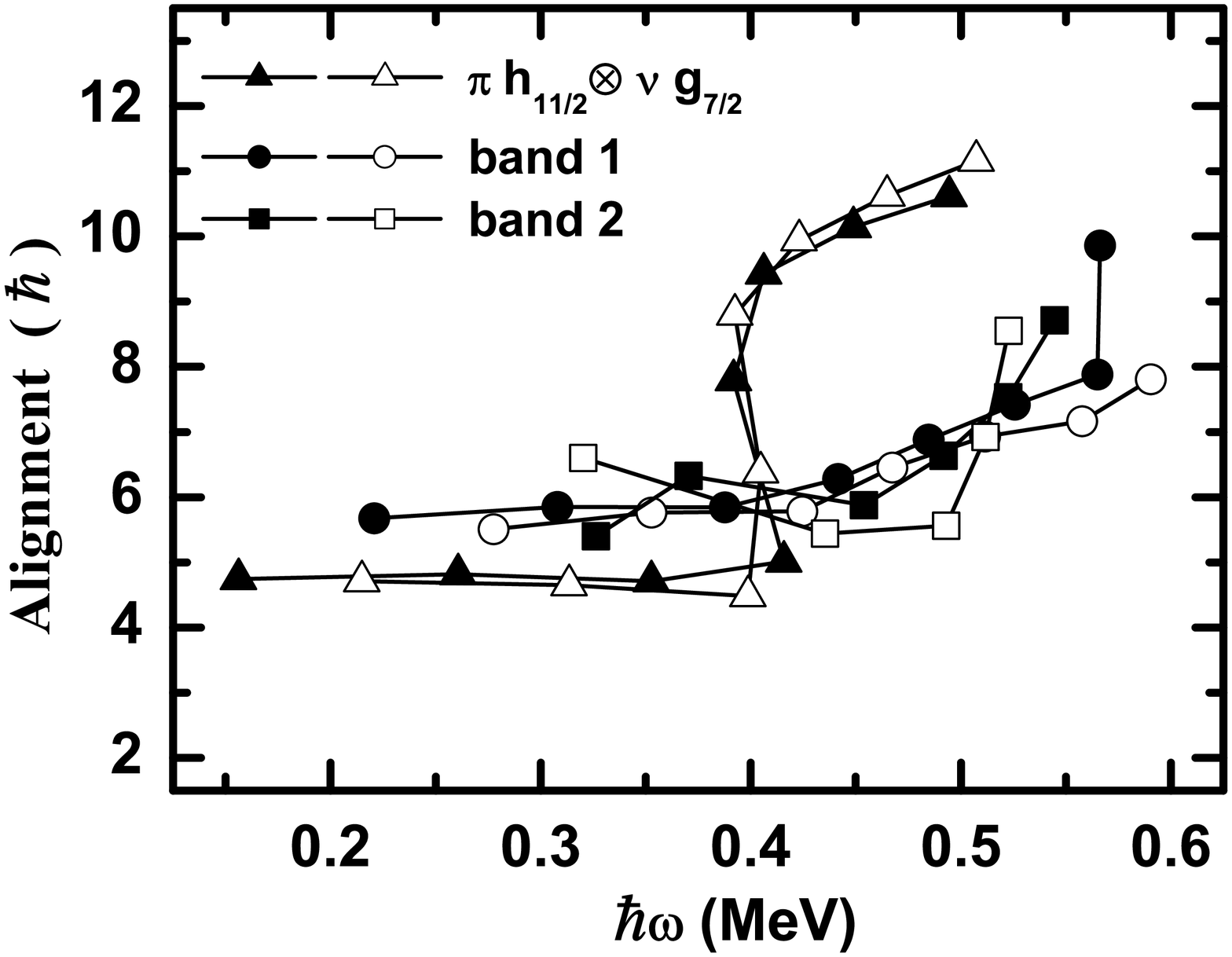}% Here is how to import EPS art
\caption{\label{fig:epsart} Rotational alignments of bands in
$^{126}$Cs. Harris parameters ($J_{0}$=17.0,
$J_{1}$=25.8)\cite{Liang90} were used to subtract the angular
momentum of the core.}
\end{figure}

\begin{figure}[h!]
\includegraphics[bb=15 15 295 237,scale=0.80]{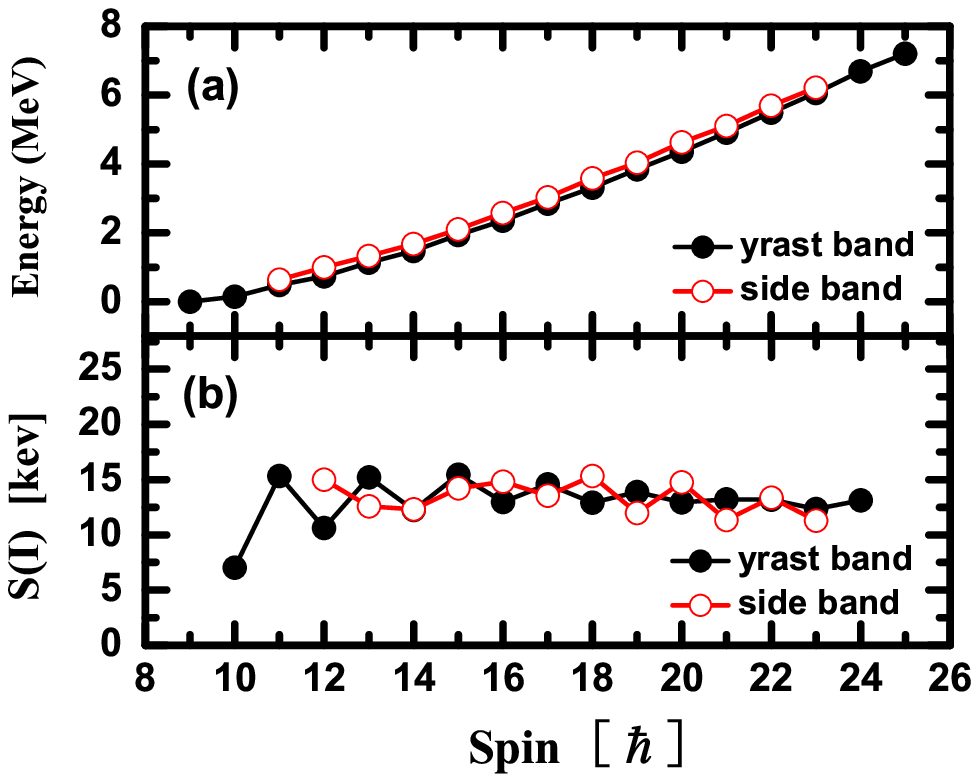}% Here is how to import EPS art
\caption{\label{Fig. :epsart} Excitation energy vs. spin (upper
panel) and S(I) values vs. spin for the doublet bands in
$^{126}$Cs.}
\end{figure}

\begin{figure}[h!]
\includegraphics[bb=35 215 566 636,scale=0.75]{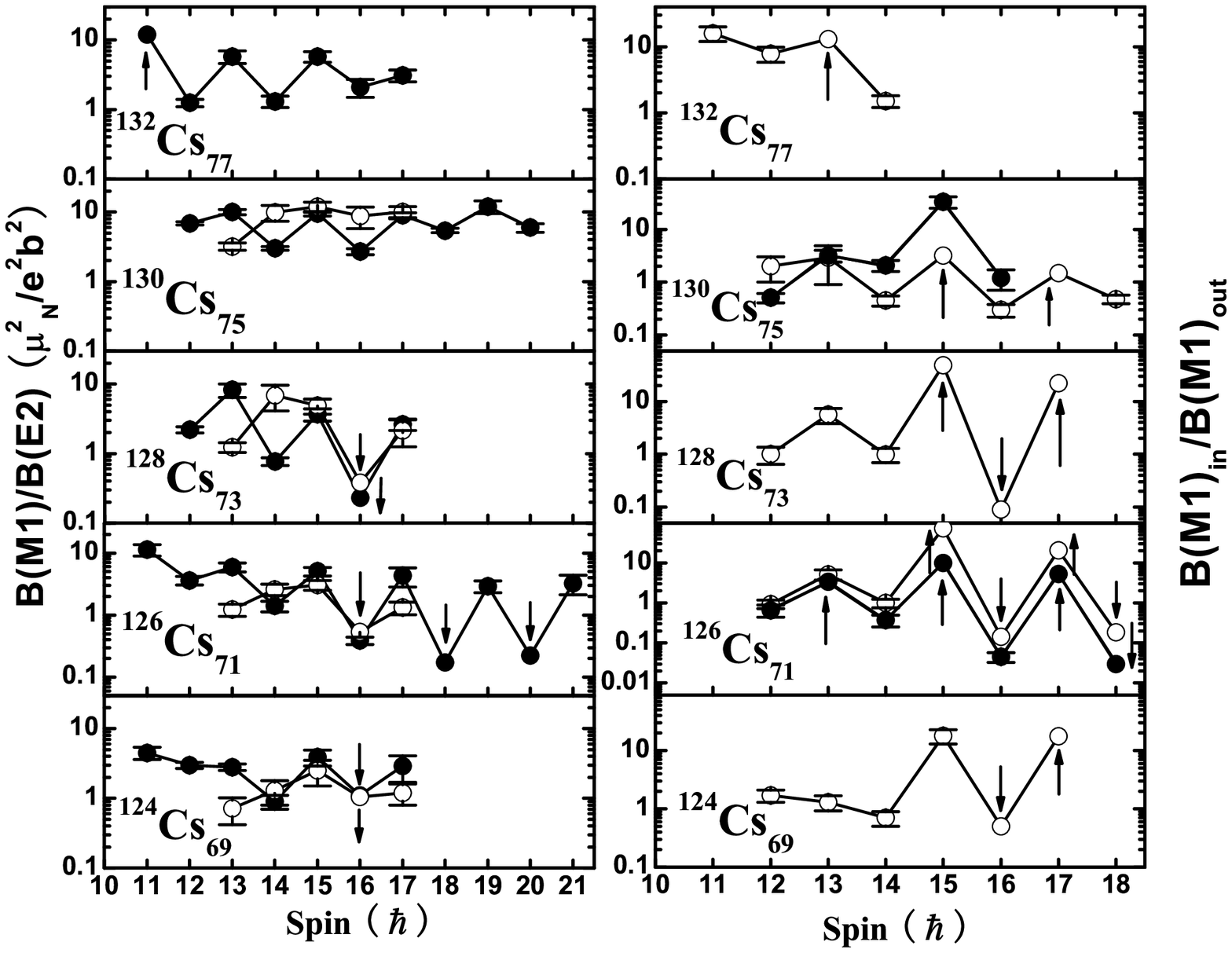}% Here is how to import EPS art
\caption{\label{fig:wide}$B(M1)/B(E2)$ and
$B(M1)_{in}/B(M1)_{out}$ versus spin for yrast and side bands in
$^{126}$Cs to compare with those in $^{124}$Cs\cite{AG01},
$^{128}$Cs\cite{Koike03}, $^{130}$Cs\cite{AJ05},
$^{132}$Cs\cite{GR03}. Points with up (down) arrows represent
lower (upper) limits.}
\end{figure}


\normalsize

\begin{thebibliography}{100}
\bibitem{FM97} S. Frauendorf and J. Meng, Nucl. Phys. {\bf{A617}},131
(1997)
\bibitem{Koike01}T. Koike, K. Starosta, C.J. Chiara, D.B. Fossan,
and D.R. LaFosse, Phys. Rev. {\bf{C63}}, 061304(R) (2001)
\bibitem{Starosta01}K. Starosta \emph{et al.,}
Phys. Rev. Lett. {\bf{86}}, 971 (2001)
\bibitem{GR03} G. Rainovski \emph{et al.,} Phys. Rev. {\bf{C67}}, 024318 (2003)
\bibitem{Bark01}  R.A. Bark, A.M. Baxter, A.P. Byrne, G.D. Dracoulis,
 T. kibedi, T.R. McGoram and S.M. Mullins, Nucl. Phys. {\bf{A691}}, 577 (2001)
\bibitem{Koike03} T. Koike, K. Starosta, C.J. Chiara, D.B. Fossan,
and D.R. LaFosse, Phys. Rev. {\bf{C67}}, 044319 (2003)
\bibitem{AJ05} A.J. Simons \emph{et al.,} J. Phys. {\bf G31}, 541 (2005)
\bibitem{AG01} A. Gizon \emph{et al.,} Nucl. Phys. {\bf{A694}}, 63 (2001)
\bibitem{LiXF02} X. Li \emph{et al.,} Chin. Phys. Lett. {\bf{19}}, 1779 (2002)
\bibitem{TK93} T. Komatsubara \emph{et al.,} Nucl. Phys. {\bf{A557}}, 419c
(1993)
\bibitem{MP96} M. Piiparinen \emph{et al.,} Nucl. Phys. {\bf{A605}}, 268
(1996)
\bibitem{WANG06} S. Wang \emph{et al.,} J. Phys. {\bf G32}, 283 (2006)
\bibitem{YLiu98} Y. Liu, J. Lu, Y. Ma, S. Zhou, and H. Zheng,
Phys. Rev. C {\bf{58}}, 1849 (1998)
\bibitem{LiXF03} X. Li \emph{et al.,} Eur. Phys. J. A {\bf{17}}, 523 (2003)
\bibitem{Liang90} Y. Liang \emph{et al.,} Phys. Rev. {\bf{C42}}, 890 (1990)
\bibitem{Koike02} T. Koike \emph{et al.,} AIP Conf. Proc. {\bf 656}, 160
(2002)
\end{thebibliography}
\end{document}